\documentclass[twocolumn,showpacs,preprintnumbers,prb,amsmath,amssymb]{revtex4}

\usepackage{graphicx}   
\usepackage{bm}
\usepackage{epsfig}

\newcommand{\SrNi}{SrFe$_{2-x}$Ni$_x$As$_2$ }
\newcommand{\SrPt}{SrFe$_{2-x}$Pt$_x$As$_2$ }

\newcommand{\BaCo}{BaFe$_{2-x}$Co$_x$As$_2$ }
\newcommand{\BaNi}{BaFe$_{2-x}$Ni$_x$As$_2$}
\newcommand{\BaPd}{BaFe$_{2-x}$Pd$_x$As$_2$ }

\newcommand{\BaPt}{BaFe$_{2-x}$Pt$_x$As$_2$ }

\newcommand{\SrK}{Sr$_{1-x}$K$_x$Fe$_2$As$_2$ }
\newcommand{\BaK}{Ba$_{1-x}$K$_x$Fe$_2$As$_2$ }
\newcommand{\SrFeAs}{SrFe$_2$As$_2$ }
\newcommand{\BaFeAs}{BaFe$_2$As$_2$ }
\newcommand{\ie}{{\it i.e.}}

\begin{document}

\title{Superconductivity and magnetism in platinum-substituted SrFe$_2$As$_2$ single crystals}

\author{Kevin Kirshenbaum}
\author{Shanta R. Saha}
\author{Tyler Drye}
\author{Johnpierre Paglione}
\email{paglione@umd.edu}
\affiliation{Center for Nanophysics and Advanced Materials, Department of Physics, University of Maryland, College Park, MD 20742}

\date{\today}

\begin{abstract}

Single crystals of SrFe$_{2-x}$Pt$_x$As$_2$ (0 $\leq$ x $\leq$ 0.36)
were grown using the self flux solution method and
characterized using x-ray crystallography, electrical transport,
magnetic susceptibility, and specific heat measurements.  The
magnetic/structural transition is suppressed with increasing Pt
concentration, with superconductivity seen over the range 0.08 $\leq$
x $\leq$ 0.36 with a maximum transition temperature $T_c$ of 16 K at
$x$ = 0.16. The shape of the phase diagram and the changes to the
lattice parameters are similar to the effects of other group
VIII elements Ni and Pd, however the higher transition temperature
and extended range of superconductivity suggest some complexity beyond the
simple electron counting picture that has been discussed thus far.

\end{abstract}


\maketitle

\section{Introduction}

The discovery of superconductivity in iron-based compounds has led
to a renewed interest in the field of superconducting materials
research\cite{Kamihara3296}. This new class of superconductors
offers a new perspective from which the phenomenon of
superconductivity can be studied~\cite{Reviews}.  The
similarities between the phase diagrams of these systems, as well as their likeness to 
those found in cuprate and heavy-fermion systems \cite{ButchAJP76}, makes the study of these phase diagrams 
not only interesting in the context of pnictide superconductors but even more generally to the
study of unconventional superconductivity.

The family of iron-based superconductors is now quite large.
After the initial discovery of superconductivity in LaFeAsO$_{1-x}$F$_x$ there
have been several other classes discovered, colloquially named after
the stoichiometry of the parent compounds, such as ``111" for
LiFeAs\cite{Pitcher5918}, ``11" for FeSe and
FeTe~\cite{McQueen014522}, and the largest so far ``22426" for
Fe$_2$As$_2$Sr$_4$V$_2$O$_6$ \cite{Zhu220512}.  Of the different families of
pnictide superconductors discovered, the largest single crystals
have been grown in the AFe$_2$As$_2$ phase (A = alkaline earth), known as the ``122"
phase, which has the ThCr$_2$Si$_2$ crystal structure.  In this
phase alkali substitution on the alkaline earth site professes the
highest superconducting transition temperatures with 38~K in \BaK
\cite{Rotter107006} and 37~K in \SrK \cite{Sasmal107007, Chen3403}.
It has also been found that substituting Co, Ni, Pd, Ru, Rh, and Ir in \SrFeAs suppresses the magnetic (structural) transition and induces superconductivity \cite{Han024506, Schnelle214516, LeitheJasper207004, Ni024511, Saha224519} with a maximum $T_c$ between 10 and 20 K while Mn and Cr doping suppress the magnetic transition without the onset of superconductivity \cite{Sefat224524, Canfield060501, Kasinathan025023, Kim10040659}. Superconductivity has also been found in isovalent chemical substitution, substituting Ru for Fe, or P for As, in \BaFeAs \cite{Paulraj09022728, Qi09034967, Jiang382203, Kasahara09054427}. However, chemical substitution is not the only way to induce superconductivity
in the iron-based superconductors; the suppression of magnetic order
and the onset of superconductivity have also been shown to occur in
the parent compounds under pressure\cite{Alireza012208, Kumar184516}
and lattice strain\cite{Saha037005}.

Recently superconductivity has been discovered with a transition temperature $T_c$ of 23~K in single-crystal samples  of \BaFeAs with 5\% Fe substituted by Pt (Ref.~\onlinecite{SahaJPCM072204}), and later shown to exist in a range of Pt substitutions in polycrystalline samples of the \BaPt\ series\cite{Zhu104525}.
The transition temperatures in \BaPt\ are comparable to those of other transition metal substituted \BaFeAs systems, but occur over a substitution range that extends slightly beyond that of the related Group VIII substitution systems \BaNi and \BaPd (Ref.~\onlinecite{Ni024511}).  
Here we report on the phase diagram of superconducting and magnetic phases in the Pt-substituted series \SrPt. We have investigated the progression of physical properties of single crystal samples of \SrPt\ by measuring x-ray diffraction, electrical resistivity, magnetic susceptibility and specific heat measurements to characterize the evolution ground states in this system. 
We find a superconducting phase that exists over an extended range of Pt concentrations, with a maximum transition temperature of 16~K. The overall properties of Pt-substituted samples evolve similar to those of several other transition metal substitution series in the 122 materials, but with an intermediate value of the superconducting transition temperature as compared to other systems.

\section{Experimental}

Single crystals of SrFe$_{2-x}$Pt$_x$As$_2$ were prepared using the FeAs self flux method described in detail elsewhere\cite{Saha224519}.  Starting materials (Sr (99.95\% purity), Fe (99.999\%), Pt (99.99\%), and As (99.99\%)) were mixed and placed in alumina crucibles, heated to 1150 $^{\circ}$C and cooled at a rate of 2 $^{\circ}$C/hr to a temperature below the FeAs melting point after which they were quenched in the furnace.  Crystal specimens were mechanically separated from the excess flux for measurements  and cut into bar shape with typical dimensions ($1.50 \times 1.50 \times 0.100$) mm$^3$.  The samples were characterized with a Bruker D8 diffractometer using Cu K$\alpha$ x-rays at room temperature. 

Chemical analysis was obtained using electron probe microanalysis with wavelength dispersive spectroscopy (WDS), the results of which gave the Pt content and showed proper stoichiometry for each crystal shown. Resistivity measurements were performed using the standard 4-probe ac method, via gold wire / silver paint contacts with typical contact resistances of $\sim 1~\Omega$ at room temperature, and using 0.1~mA excitation currents at low temperatures. Magnetic susceptibility was measured using a commercial SQUID magnetometer, and specific heat was measured using the thermal relaxation method.

\begin{figure}[!t]\centering
  \resizebox{8cm}{!}{
    \includegraphics[width=8.5cm]{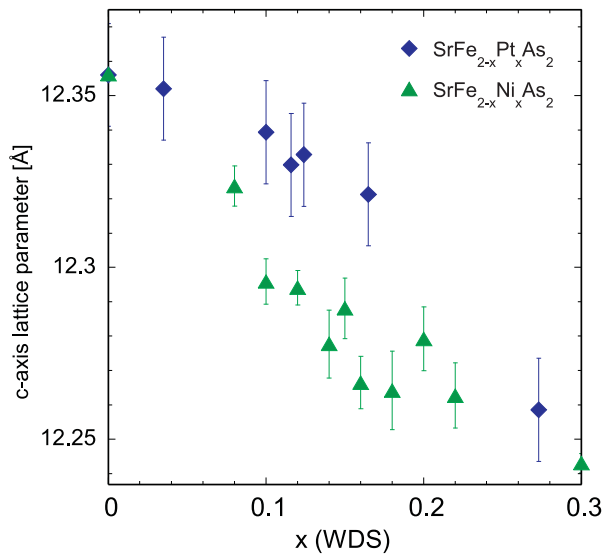}}
    \caption{\label{fig1} Variation of the $c$-axis lattice parameters with Pt concentration in single-crystal samples of SrFe$_{2-x}$Pt$_x$As$_2$ compared with that from powder data of \SrNi\ from Ref.~\onlinecite{Saha224519}. The $c$-axis length contracts with increasing Pt content, consistent with previous chemical substitution studies. Error bars were obtained by finding the lattice constant at the positions of the half maximum points on the x-ray diffraction peaks for each sample.}
\end{figure}

\section{Results and Discussion}

\subsection{Chemical and Structural Analysis}

In this study crystals were grown with as much as 18$\%$ Pt substituted for Fe (\ie, $x=0.36$).
The platinum content of single-crystal samples was measured using WDS, with analysis 
performed at several points along each crystal with spot sizes $\sim$~1 $\mu$m$^2$. While the Pt content within each single crystal specimen was found to be uniform, a dispersion of Pt content was found 
within some of the batches.  Because of this, each sample used for measurements presented in this work was measured individially by WDS to determine the chemical composition, 
rather than the more common practice of analyzing a few crystals
from each nominal stoichiometry and using the representative value for all samples from the batch. The results of such analysis reveal a slight gap in the range of Pt concentrations, approximately between $0.18 \leq x \leq 0.26$, despite repeated attempts at crystal growth targeted for this range. Future work will address the reason for these difficulties.

\begin{figure}[!t]\centering
  \resizebox{8cm}{!}{
    \includegraphics[width=8.5cm]{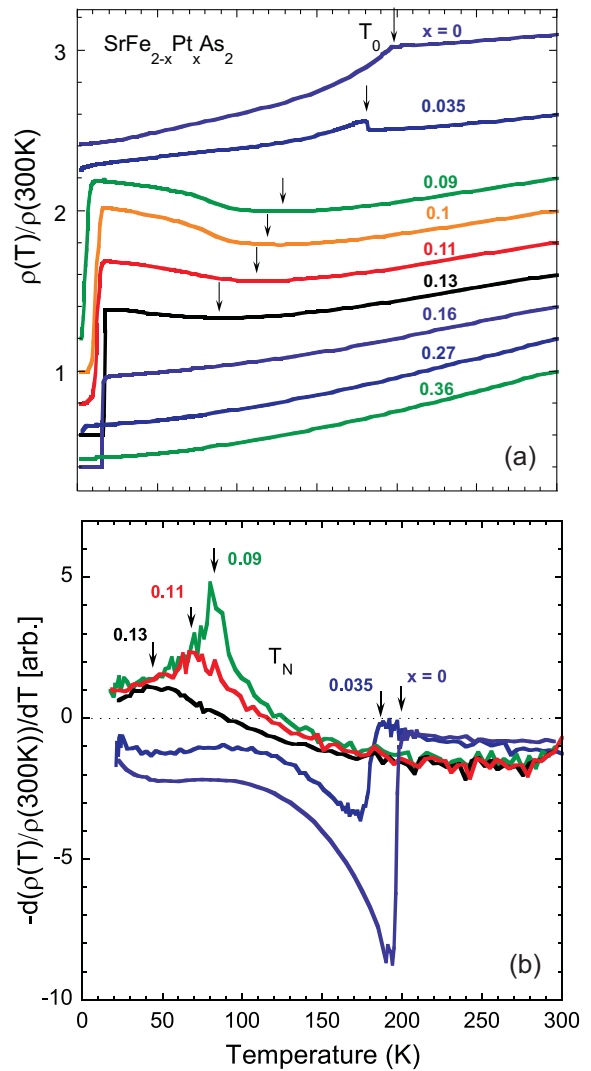}}
    \caption{\label{fig2} (a): electrical resistivity of single-crystal samples of SrFe$_{2-x}$Pt$_x$As$_2$~normalized to 300~K (data are shifted vertically for clarity for all concentrations except $x=0.36$). The structural transition $T_0$, identified by the local mimina in $\rho(T)$, is suppressed with increasing Pt substitution, as indicated by the arrows.  (b): negative of the temperature derivative of the normalized resistivity data for characteristic samples from part (a). The antiferromagnetic transition $T_N$ is identified at the local maxima, while the dashed line indicates the zero crossing of $-d\rho/dT$ corresponding to local minima in $\rho(T)$. At optimal doping, $x$ = 0.16, $T_0$ and $T_N$ are no longer discernible and the maximum $T_c$ for the system is found to be 16~K. Neither magnetic order nor superconductivity is observable at $x$ = 0.36, the highest concentration sample measured in this study.} 
\end{figure}

Single crystal x-ray diffraction was used to obtain the $c$-axis
lattice parameters for crystalline samples across the series. As shown in Fig.~\ref{fig1}, 
the unit cell contracts in the $c$-axis direction in this system as it does for other transition metal substitutions in the SrFe$_2$As$_2$ system\cite{Saha224519, Schnelle214516,
LeitheJasper207004, Han024506}. Notably, the progression of the $c$-axis
lattice parameter follows almost 	exactly that of the Ni-doped series \SrNi.
Since Pt and Ni have different covalent radii (1.36 \AA\ for Pt and 1.24 \AA\ for Ni) it's surprising that within the experimental accuracy the absolute values
of lattice size in the Pt system approximate those seen in the Ni
system, decreasing to 12.25 \AA\ for $x$ = 0.28 (see fig. 1). This is suggestive
of the minimal effect of the nature of substituent atoms on the unit cell evolution, 
likely dominated by modifications to the electronic structure of the FeAs layers at these lower concentration levels.

\subsection{Electronic Transport}

Fig.~\ref{fig2} presents the temperature dependence of electrical
resistivity, $\rho(T)$, in SrFe$_{2-x}$Pt$_x$As$_2$ normalized to
the room temperature value as well as the derivative of this normalized resistivity, $-d(\rho(T)/\rho(300~K))/dT$. This choice of normalization is done to remove the error contribution coming 
from the geometric factor due to unobservable internal cracks or exfoliation, 
with data in Fig.~\ref{fig2} shifted vertically for clarity.  
At high temperatures, the resistivity is fairly 
linear in temperature and with the same slope for all concentrations of Pt,
indicating that the scattering at high temperatures has very little
electronic dependence and is likely dominated by phonons or some other
mechanism that is independent of Pt substitution. In the parent
compound \SrFeAs, an onset of antiferromagnetic order at $T_N$ as well as a change from
tetragonal to orthorhombic crystal structure at $T_0$ both occur at a
characteristic temperature of 200~K, which manifests itself as a sharp drop 
in the resistivity upon decreasing temperature. With substitution of as little as 2.0\% Pt in place of Fe, this drop through the transition transforms into an upturn below $T_0$.  The evolution of the character of the transition in resistivity, from downturn to upturn, can be explained in terms of a shift in the balance between the loss of inelastic (magnetic) scattering due to the onset of magnetic order and the change in carrier concentration associated with the transition at $T_0$, as observed in several other FeAs-based systems.\cite{Saha224519} Upon further Pt substitution, the transition then appears as a broader minimum in the resistivity which becomes less discernible as a sharp feature \cite{Ni024511}.

\begin{figure}[!t]\centering
  \resizebox{8cm}{!}{
    \includegraphics[width=8cm]{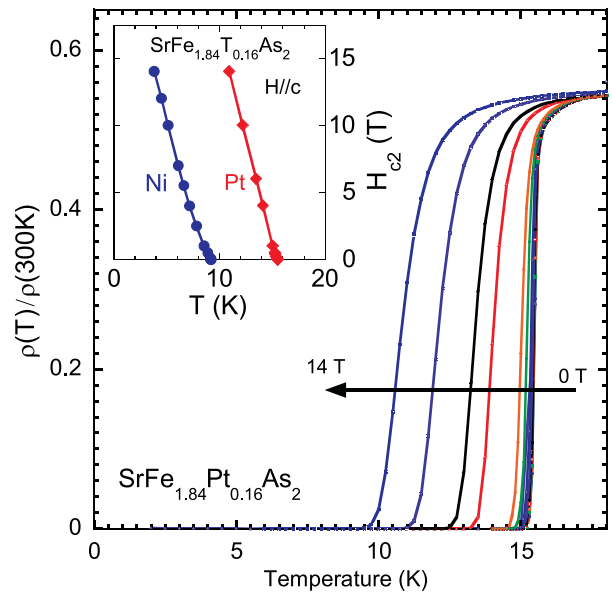}}
    \caption{\label{fig3} Magnetic field suppression of the resistive superconducting transition in optimally doped SrFe$_{1.84}$Pt$_{0.16}$As$_2$ for fields up to 14~T applied parallel to the $c$-axis. The resistivity is normalized to the room temperature value.
    Inset: superconducting upper critical field $H_{c2}$(T), defined at the 50\% point of the normal state resistivity for each field in the main panel, plotted with that in SrFe$_{2-x}$Ni$_x$As$_2$ from Ref.~\onlinecite{Butch024518}.}
\end{figure}

The position of $T_0$, marked by arrows in part (a) of Fig.~\ref{fig2}, is identified with the onset of the sharp drop for $x\leq 0.035$, and the position of the minimum in resistivity for $x\geq 0.09$.In the related system \BaCo, the magnetic and structural transitions that occur simultaneously at 200~K in the parent compound has been shown to split into two separate transition temperatures upon Co doping \cite{Chu014506}. As done in Ref.~\onlinecite{Chu014506}, we examine the derivative of resistivity, $d\rho/dT$, in order to identify the splitting of the structural ($T_0$) and magnetic ($T_N$) transitions upon Pt substitution. Shown in Fig.~\ref{fig2}b, $-d\rho/dT$ indeed exhibits a peak whose transition temperature value differs from the minimum in $\rho(T)$ used to identify $T_0$ for $x\geq 0.09$. Given the resolution of our data, we are unable to resolve two distinct features in the derivative plot as done in the study by Chu {\it et al.},\cite{Chu014506} and therefore define
$T_N$ at the position of the maximum in $-d\rho/dT$ as indicated by arrows in Fig.~\ref{fig2}b). The peak widths are used to define the uncertainty in $T_N$ values associated with this assignment (see phase diagram in Fig.~\ref{fig6}).

At $x=0.09$, superconductivity becomes visible in the resistivity, where it coexists with the magnetically ordered phase. With further substitution the magnetism is gradually suppressed and vanishes near $x$ = 0.16 (hereafter called optimal doping), where the maximum $T_c$ of 16 K is observed. For $x$ = 0.36 both magnetic order and superconductivity are suppressed and normal metallic behavior appears to be recovered.

\begin{figure}[!t]\centering
  \resizebox{8cm}{!}{
    \includegraphics[width=8cm]{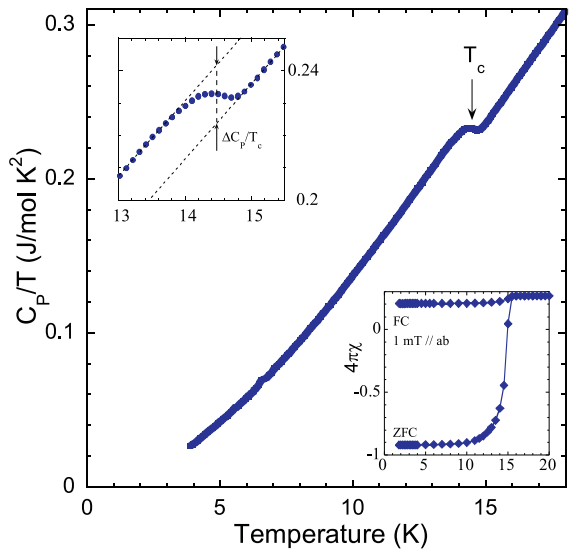}}
    \caption{\label{fig4} Specific heat capacity of optimally doped 
    SrFe$_{1.84}$Pt$_{0.16}$As$_2$, with arrow indicating superconducting transition temperature. Upper inset: zoom of $C_p(T)/T$ near the superconducting transition, with lines indicating fits to the data above and below $T_c$ and arrows indicating the position where $\Delta C_p/T_c$ is determined.  Lower inset: volume magnetic susceptibility of SrFe$_{1.84}$Pt$_{0.16}$As$_2$ measured in 1~mT field applied parallel to the crystallographic basal plane, following both zero-field-cooled (ZFC) and field cooled (FC) conditions. The sharpness of the transition, with onset near 16~K, is identified by the complete saturation of Meissner screening by 10~K. The bulk nature of the transition is shown by the achievement of nearly full screening fraction of $4\pi\chi$.}
\end{figure}

Fig.~\ref{fig3} presents the evolution of the superconducting
transition in resistivity for the optimal doping $x$ = 0.16 in
a magnetic field $H$. With the increase of $H$ there is a
slight broadening of the transition, with the narrowest transition width of $\sim 0.5$~K for $H$=0 
increasing to the broadest width of $\sim 2.5$~K at $H$=14 T.  
The inset of Fig.~\ref{fig3} shows the temperature
dependence of the superconducting upper critical field $H_{c2}$,
along with data for optimally doped \SrNi\ from
Ref.~\onlinecite{Saha224519} for comparison. (The value
of $H_{c2}$ is defined by the field where resistivity drops to 50\% of the normal state resistivity.) Using the Werthamer-Helfand-Hohenberg (WHH) formula\cite{WHH},
$H_{c2}$(0) is calculated to be 34~T for SrFe$_{1.84}$Pt$_{0.16}$As$_2$ and was confirmed with $C_p(T)$ in field measurements (not shown).  While this is much higher than that
found in the Ni system the value of the slope $dH_{c2}/dT$ near $T_c$ for Pt of -3.16 T/K, is comparable to that of several other transition metals substitutions in the 122 systems \cite{Butch024518}, indicating a similar nature of the $H_{c2}(T)$ transition in all of these systems.

\subsection{Specific Heat and Magnetic Susceptibility}

\begin{figure}[!t]\centering
  \resizebox{8cm}{!}{
    \includegraphics[width=8cm]{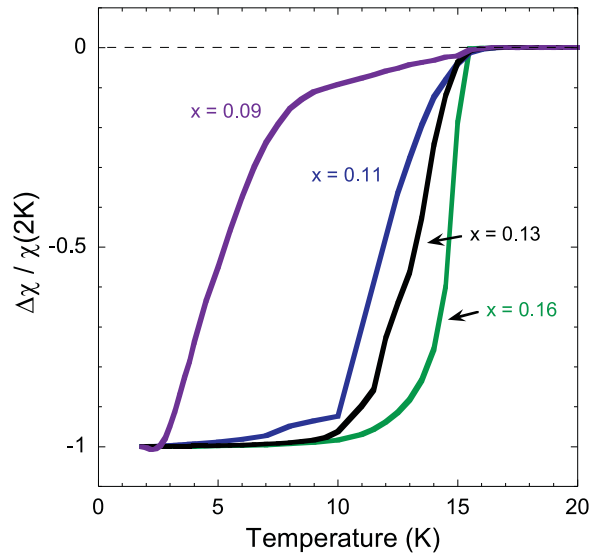}}
    \caption{\label{fig5} Change in magnetic susceptibility ($\Delta\chi\equiv\chi(20~K) - \chi(T)$) of SrFe$_{2-x}$Pt$_{x}$As$_2$ normalized to the susceptibility value at 2~K.  Samples were measured in 1~mT field applied parallel to the crystallographic basal plane for zero-field-cooled (ZFC) and field cooled (FC) conditions.}
\end{figure}

To verify the bulk thermodynamic nature of the superconducting transition,
specific heat measurements were performed using an optimally doped
sample, SrFe$_{1.84}$Pt$_{0.16}$As$_2$, with resistive superconducting transition at 16~K.  As shown in Fig.~\ref{fig4}, the superconducting transition is readily identified as an abrupt shift in the smooth $C_p(T)/T$ curve below 15 K, consistent with the value of $T_c$ observed in resistivity. Because the values of both $T_c$ and $H_{c2}$ in this system are high, a reliable estimate of the normal state electronic specific heat $\gamma$ is difficult due to the dominant phonon contribution. 
For this reason we have determined $\Delta C_p/T_c$ rather than the more traditional quantity $\Delta C_p/\gamma T_c$ as shown in the inset of Fig.~\ref{fig4}. An
isentropic construction, as done previously \cite{SahaJPCM072204,Budko220516}, gives a value of $\Delta C_p/T_c \simeq 17$~mJ/mol~K. Assuming the BCS weak-coupling approximation $\Delta C_p/\gamma T_c$=1.43 the value of $\gamma$ is estimated to be $\sim 12$ mJ/mol K$^2$, comparable to that found in other transition
metal-doped \BaFeAs superconductors \cite{SahaJPCM072204}. In Ref.~\onlinecite{Budko220516}, the values of $\Delta C_p/T_c$ measured for K, Co, Ni, Rh, and Pd-doped \BaFeAs\ superconductors have been shown to scale with $T_c$, regardless of the value of $T_c$ or the relative substituent concentration (\ie, either under- or overdoped) with respect to maximum $T_c$. Similar to \BaPt \cite{SahaJPCM072204}, the corresponding value of $\Delta C_p/T_c \simeq$ 17 mJ/mol~K taken at 14.5 K for SrFe$_{1.84}$Pt$_{0.16}$As$_2$ falls in line with this
trend, expanding this interesting relation to include another
$5d$-transition metal-doped Sr analogue. Although the reason for
this relation is unknown, it indicates some universal effect in
pnictide superconductivity. 

The bulk nature of the superconducting transition is also indicated through the observation of Meissner screening in low-field magnetic 
susceptibility measurements. Because of the small size of available samples, normalized data is presented in order to provide for a more useful comparison. Fig.~\ref{fig5} presents the temperature dependence of the change in susceptibility from the normal state at 20~K normalized to the value at 2~K, $\Delta\chi(T)/\chi(2~K)$ where $\Delta\chi = \chi(20~K) - \chi(T)$. Data were measured in a magnetic field of 1~mT applied parallel to the $ab$-plane both in zero field cooled (ZFC) and field cooled (FC) conditions. In the ZFC measurement the samples were initially cooled down to 1.8 K in zero magnetic field and then 1~mT DC magnetic field was applied. The saturation of magnetic susceptibility indicates full Meissner screening and hence a nearly full superconducting volume fraction. The difference between FC and ZFC curves is due to trapped flux remaining in the sample as it is cooled through the superconducting transition in field. 

As shown, the optimal doped sample $x=0.16$ shows the sharpest and highest $T_c$, consistent with the specific heat and resistivity data, while the major drop in $\Delta\chi$ is suppressed in temperature for lower Pt concentrations.  It is apparent that the $x=0.09$ data shows two transitions, including a mild drop in $\Delta\chi$ at 15.5~K and a much sharper drop at 8~K, indicating a sample with mixed phase. This is likely due to the inclusion of a crystallite with higher Pt concentration in the collection of several small pieces required to increase signal to noise to a measurable level. No magnetic transition was
observed in SrFe$_{1.84}$Pt$_{0.16}$As$_2$ in the normal state above $T_c$ (not shown), consistent with the absence of a signature of the transition in resistivity data.

\begin{figure}[!t]\centering
    \includegraphics[width = 8cm]{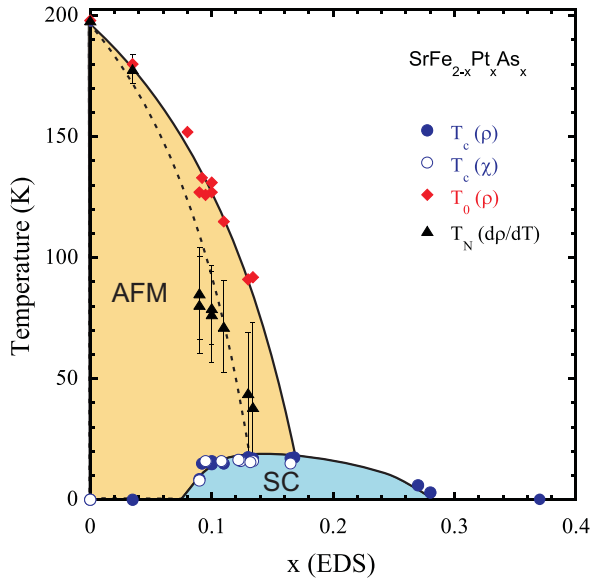}
    \caption{\label{fig6} Phase diagram of the \SrPt system.  Magnetic (triangles), structural (diamonds) and superconducting (circles) transition temperatures are obtained from resistivity (solid symbols) and magnetic susceptibility (open symbols). Solid lines are added as guides for the structural and superconducting transitions. The error bars for the magnetic transition temperatures represent the width of the peak in the $d\rho/dT$ data (see text).  }
\end{figure}

\subsection{Phase Diagram}

Combining the results of transport and magnetic susceptibility we present the phase diagram of Pt substitution for the \SrPt\ series in Fig.~\ref{fig6}. The values of $T_0$ and $T_N$ have been explained in a previous section and $T_c$ is defined at the onset of the resistive transition as determined by the intercept of the extrapolation of the normal state data and the mid-transition data. The values of $T_c$ obtained from resistivity match with those signifying the onset in susceptibility for all samples measured (in the range $0.09\leq x \leq 0.16$), as well as the transition measured in specific heat.

The phase diagram for \SrPt\ system is similar to other electron-doped members of the 122 family of iron-based superconductors \cite{Reviews}. 
The antiferromagnetic ordering is suppressed with increasing transition metal substitution, and a superconducting phase is stabilized at some finite concentration.  Using these definitions for $T_0$ and $T_N$ we observe a splitting of the magnetic and structural transitions. Though generally referred to as a superconducting ``dome'', the shape of the superconducting region of the phase diagrams for \SrPt\ and \SrNi\ are more step-like on the underdoped side than they are dome-like.  In the case of \SrPt, $T_c$ is already close to the maximum value of 16~K at the lowest Pt concentration where superconductivity onsets ($x \simeq 0.08$).  

This behavior is also observed in both \SrNi\ (Ref.~\onlinecite{Saha224519}) and CaFe$_{2-x}$Ni$_x$As$_2$ (Ref.~\onlinecite{Kumar09082255}), as well as a recent study of single-crystalline SrFe$_{2-x}$Co$_x$As$_2$ samples grown using Sn flux (Ref.~\onlinecite{Kim10040659}). The SrFe$_{2-x}$Co$_x$As$_2$ study also found that the superconducting dome extends further into the underdoped region than previous work
\cite{LeitheJasper207004}. In \SrPt, the superconducting dome is similarly extended significantly beyond that seen in the Ni-substituted system, with a larger range of concentrations exhibiting superconductivity.  It also appears that the superconducting dome may be better described as a plateau.  This
effect does not appear to be universal throughout all 122 systems, though, 
as transition metal substitution in BaFe$_2$As$_2$ generally shows a gradual, less abrupt onset of the superconducting dome as a function of doping \cite{Ni024511, Ni214515, Canfield060501}. It may be particular to Sr-based 122 systems, but more work is required to properly compare high-quality specimens grown under similar conditions and measured using similar techniques before grander conclusions can be made. 

Nevertheless, it is apparent that the antiferromagnetic and superconducting phases share a wide range of coexistence between the $T_c$ onset and optimal doping; future studies determining the extent and nature of the coexistence of these two phases will be useful for comparison to other 122 superconducting systems. Finally, it is an important question as to why superconductivity in Sr-based 122 materials substituted with Ni\cite{Saha224519}, Pd\cite {Han024506} and now Pt (this work) tend to show lower maximum values of $T_c$ as compared to their Ba-based counterparts\cite{SahaJPCM072204}. Ongoing work\cite{pairbreaking} suggests that pair-breaking effects may have a role in limiting the transition temperature of some members of the 122 family of iron-pnictide superconductors.

\section{Summary}

In conclusion, single crystals of Pt-substituted SrFe$_2$As$_2$ were
grown and characterized with up to 18$\%$ substitution and a phase
diagram is determined. Similar to other transition metal-substituted
systems, the magnetic and structural ordering seen at 200~K in
\SrFeAs is suppressed with chemical substitution as in several other 122 iron pnictide materials, with a superconducting phase reaching a maximum transition temperature
of $T_c$~=~16~K at an optimal doping $x=0.16$ where the magnetic and
structural transitions are suppressed. Future work will address the reasons for variations in maximum $T_c$ values found in SrFe$_2$As$_2$ substituted with Group VIII transition metal elements.

\begin{acknowledgments}
The authors acknowledge N.~P.~Butch and R.~L.~Greene for useful discussions. This
work was partially supported by an AFOSR-MURI Grant (FA9550-09-1-0603) and an NSF-CAREER Grant (DMR-0952716).
\end{acknowledgments}


\end{document}